\documentclass[preprint,nofootinbib]{revtex4-1}     
\usepackage{graphicx}
\usepackage{amssymb}
\usepackage{amsmath}

\begin{document}

\title{Topological Supersymmetry Breaking as the Origin of the Butterfly Effect
}


\author{Igor V. Ovchinnikov}



\affiliation{Electrical Engineering Department, 
            University of California at Los Angeles,             Los Angeles, CA 90095 USA}
\email{igor.vlad.ovchinnikov@gmail.com} 

\begin{abstract}
Previously, there existed no clear explanation why chaotic dynamics is always accompanied by the infinitely long memory of perturbations (and/or initial conditions) known as the butterfly effect (BE). In this paper, it is shown that within the recently proposed approximation-free supersymmetric theory of stochastic (partial) differential equations (SDE), the BE is a derivable consequence of (stochastic) chaos, a rigorous definition of which is the spontaneous breakdown of topological supersymmetry that all SDEs possess. It is also discussed that the concept of ergodicy must be refined under the condition of the spontaneous breakdown of pseudo-time-reversal symmetry when the model has "physical" states of multiple eigenvalues that survive the physical limit of the infinitely long temporal propagation. 
\end{abstract}
\maketitle

\section{Introduction}

Dynamical chaos is a very fundamental phenomenon found originally by Poincar\'e \cite{Rue14} and rediscovered numerically later by Lorenz and others. \cite{Mot14,Shep14} Many important theoretical insights on chaotic dynamics have already been provided.\cite{Has03} Nevertheless, some of the most fundamental questions remained unanswered. For example, there was no rigorous generalization of chaos for stochastic models. Furthermore, there existed no convincing explanation why chaotic dynamics is always accompanied by its most definitive property of the infinitely long dynamical memory of perturbations and/or initial conditions known as the butterfly effect (BE).  

A new and interesting side of the story of chaotic dynamics has been recently found within the  approximation-free and coordinate-free supersymmetric theory of stochastic (partial) differential equations (SDEs).\cite{Ovc13,Ovc14} This theory, that can be called the supersymmetric theory of stochastic dynamics or even simpler of stochastics (STS), came out as a result of the conjecture reported in Ref.\cite{Ovc11} that dynamical chaos, or rather one type of its stochastic generalization known as self-organized criticality, may as well be the phenomenon of the spontaneous breakdown of the topological supersymmetry that all SDEs possess and the existence of which has been known for a while now (see, \emph{e.g.}, Ref.\cite{Gaw}). 

In this paper, it is shown that the BE is a derivable property from the STS picture of stochastic chaos. The structure of the paper is as follows. In Sec. \ref{DeterministicDynamics}, deterministic continuous-time dynamics is reformulated as a family of maps that induce actions on generalized probability distributions. In Sec.\ref{StochasticDynamics}, the stochastic generalization of this picture is provided. In Sec.\ref{TopSusy}, the topological supersymmetry of continuous-time (stochastic) dynamics is discussed. The eigensystem of the generalized Fokker-Planck stochastic evolution operator is addressed in Sec.\ref{Spectrum}. In Sec.\ref{SecPartitionFunction}, it is shown why topological supersymmetry breaking is the essence of dynamical chaos. The response to external perturbations and the BE are analyzed and the concept of ergodicity is discussed in Sec.\ref{SecButterflyEffect}. In Sec.\ref{States}, a brief analysis of the structure of the ground state in the deterministic limit provides additional support to the claim that the dynamical chaos is indeed the phenomenon of the topological supersymmetry breaking.  Sec.\ref{SecConclusion} concludes the paper.

\section{Deterministic Dynamics as Pullbacks}
\label{DeterministicDynamics}

The following general class of SDEs will be considered:\footnote{The use of pathintegrals allows to generalize discussion to much wider class of models. Unlike in this paper, the noise can be of any form not necessary Gaussian white and $F$ and $e$'s must not necessarily be the temporarily local functions of $x(t)$. We will not pursue these generalizations here, however.}
\begin{eqnarray}
\dot x(t)  = F (x(t)) + (2\Theta)^{1/2}e_a(x(t))\xi^a(t) \equiv \tilde F (t) .\label{SDE}
\end{eqnarray}
Here and in the following the summation over the repeated indexes is assumed, $x\in X$ is a point from a $D$-dimensional topological manifold called phase space, $X$; $F(x) \in TX_x$ is the flow vector field from the tangent space of $X$ at point $x$, $\xi^a \in \mathbb{R}^1, a=1,2...$, are noise variables, $e_a(x) \in TX_x, a=1,2...$ is a set of vector fields on $X$, and $\Theta$ is the parameter representing the intensity or temperature of the noise. The position dependent/independent $e$'s are often called multiplicative/additive noise.

At a fixed noise configuration, the SDE in Eq.(\ref{SDE}) becomes an ordinary differential equation (ODE) with time dependent flow vector field. It defines a two-parameter family of (noise-configuration-dependent) maps of the phase space onto itself:
\begin{eqnarray}
M_{tt'}:X\to X.\label{Maps}
\end{eqnarray} 
The meaning of these maps is as follows: for each $x'\in X$ and time moment, $t'$, $x(t) = M_{tt'}(x')\in X$ is the solution of this ODE with the condition  $x(t')=x'$.

The models considered here are physical, \emph{i.e.}, they have sufficiently smooth flow vector field, $F$, and $e_a$'s so that the corresponding maps are diffeomorphisms so that they are invertible and differentiable. The maps satisfy the following equations,
\begin{eqnarray}
M_{tt} = \text{Id}_{X}, M_{t''t'}\circ M_{t't} = M_{t''t},\text{ and } M_{t't} = M_{tt'}^{-1}.\label{CompositionMaps}
\end{eqnarray}

Let us assume now that at time moment, $t'$, the model is described by a total probability function, $P(t'x)$, so that the expectation value of some function, $f(x): X\to\mathbb{R}^1$, is,
\begin{eqnarray}
\overline{f}(t') = \int_{X} f(x)P(t'x)dx^1...dx^D.
\end{eqnarray}
The same expectation value at later time moment, $t>t'$, can be given now as,
\begin{eqnarray}
\overline{f}(t) = \int_{X} f(M_{tt'}(x))P(t'x)dx^1...dx^D.\label{DetermEvolution} 
\end{eqnarray}
One example of this definition of dynamics is as follows. Consider $X=\mathbb{R}^D$, and the ODE of the following simple form, $\dot x=v$, where $v$ is some constant flow vector field. The corresponding maps are defined now as $M_{tt'}(x) = x + v(t-t')$. For the "observable" function being one of the coordinates, $f(x)=x^i(x)$, Eq.(\ref{DetermEvolution}) gives $\overline{x^i}(t) = \overline{x^i}(t') + v^i (t-t')$ as it should.

One can now make the transformation of the dummy variable of integration, $x \to x = M_{t't}(x)$, so that Eq.(\ref{DetermEvolution}) becomes:
\begin{eqnarray}
\overline{f}(t) = \int_{X} f(x) M^*_{t't}\left(P(t'x)dx^1...dx^D\right).\label{Average}
\end{eqnarray}
Here, $M^*_{t't}$ is the action or the pullback induced by the inverse map, $M^*_{t't}$, on the top differential form representing the total probability distribution:
\begin{eqnarray}
M^*_{t't}(P(t'x)dx^1...dx^D) = P\left(t'M_{t't}(x)\right) J(TM_{t't}(x))dx^1...dx^D,\label{PullbackP}
\end{eqnarray}
where $J$ is the Jacobian of the tangent map, $TM_{t't}(x): TX_{x} \to TX_{M_{tt'}(x)}$,
\begin{eqnarray}
TM_{t't}(x): dx^i \mapsto d (M_{t't}(x))^i = TM_{t't}(x)^i_k dx^k,\label{DefinitionOfTangentMap}
\end{eqnarray}
with
\begin{eqnarray}
TM_{t't}(x)^i_k = \partial (M_{t't}(x))^i/\partial x^k,\label{TangentMap}
\end{eqnarray}
being the coordinate representation of the tangent map.

\begin{figure}
\centerline{\includegraphics[height=5.0cm, width=10.5cm]{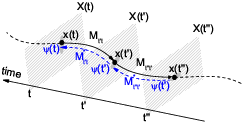} }
\caption{\label{Figure1} Stochastic differential equation (\ref{SDE}) with a fixed noise configuration is an ordinary differential equation that can be viewed as a two-parameter family of maps, $M_{t't}:X(t)\to X(t')$, between the infinite number of copies of the phase space, $X(\tau)$, for each time moment, $\tau$. At this, the forward temporal evolution of the generalized probability distributions, which are differential forms on $X$, is defined as the corresponding pullbacks induced by maps of the inverse temporal evolution: $M^*_{t't}:\Omega(X(t'))\to\Omega(X(t))$.}
\end{figure}

The conventional description of SDEs and random dynamical systems in terms of only total probability distributions (TPD's), however, is not complete as it will be discussed in Sec.\ref{SecPartitionFunction} below. In the general case and especially for chaotic dynamics, one has to consider generalized probability distributions (GPDs). In the coordinate-free setting, the GPDs are differential forms: \cite{Ovc13}
\begin{eqnarray}
\psi^{(k)} = (1/k!) \psi^{(k)}_{i_1...i_k} dx^{i_1}\wedge ... \wedge dx^{i_k} \in \Omega^k(X).\label{DefinitionOfForm}
\end{eqnarray}
Here $0\le k\le D$, $\psi^{(k)}_{i_1...i_k}(tx)$ is an antisymmetric contravariant tensor, $\wedge$ denotes wedge product, and $\Omega^k(X)$ is the space of all differential forms of degree $k$, or k-forms. The exterior algebra of $X$, $\Omega =\bigoplus_{k=0}^D \Omega^k$, is the Hilbert space of the model.

The reason why a wavefunction, $\psi^{(k)}$, can be interpreted as a GPD can be seen as follows. Consider a $k$-dimensional submanifold, or $k$-chain, $c_k$, and the number:
\begin{eqnarray}
\int_{c_k}\psi^{(k)} = p_{c_k} \in \mathbb{R}^1.\label{PartProb}
\end{eqnarray}
If one introduces local coordinates such that the $k$-chain belongs to the submanifold cut out by, $(x^{k+1},...,x^D) = (\text{Const}^{(k+1)}, ..., \text{Const}^{(D)})$, then Eq.(\ref{PartProb}) is the probability of finding variables, $(x^1,...,x^k)$, within this $k$-chain given that all the other variables are known with certainty to be equal $(\text{Const}^{(k+1)}, ..., \text{Const}^{(D)})$.

To establish the law of the temporal evolution of k-forms, one assumes that at time moment, $t'$, the model is described by, $\psi^{(k)}(t')$, so that the expectation value defined in Eq.(\ref{PartProb}) is $\int_{c_k}\psi^{(k)}(t')$. By analogy with Eq.(\ref{Average}), the same quantity at the later time moment, $t>t'$, is
\begin{eqnarray}
p_{c_k}(t) = \int_{M_{tt'}(c_k)}\psi^{(k)}(t') = \int_{c_k}M_{t't}^*\psi^{(k)}(t').\label{EvolutionOfForm}
\end{eqnarray}
Here, $M^*_{t't}: \Omega^k(X) \to\Omega^k(X)$, is the generalization of pullback from Eq.(\ref{PullbackP}) to pullbacks acting on $\Omega^{(k)}(X)$. Explicitly,\index{Pullback}
\begin{eqnarray}
M^*_{t't} \psi^{(k)}(t'x) = (1/k!)\psi^{(k)}_{i_1...i_k}(t'M_{t't}(x)) d(M_{t't}(x))^{i_k}\wedge ... \wedge d(M_{t't}(x))^{i_1},\label{PullBack}
\end{eqnarray}
where the k-form is from Eq.(\ref{DefinitionOfForm}) and $d(M_{t't}(x))^{i}$ is from Eq.(\ref{DefinitionOfTangentMap}).

Previous discussion showed that the forward temporal propagation of variables according to Eq.(\ref{SDE}) is equivalent to pullbacks that act on GPDs:
\begin{eqnarray}
\psi^{(k)}(t') \to \psi^{(k)}(t) = M^*_{t't} \psi^{(k)}(t').
\label{EvolutionOfTPD} 
\end{eqnarray}
At this, the pullbacks are induced by maps corresponding to the backward temporal propagation as compared to the temporal propagation in the original SDE. This may sound suspicious or rather as though the temporal direction of the propagation is confused. 

To clarify the situation let us switch now to the following picture used in pathintegral representations of dynamics. In this picture, there is an infinite number of copies of the phase space, $X(t)$, for each moment of time (see Fig.\ref{Figure1}). Eq.(\ref{Maps}) defines now maps from one copy of the phase space to another, $M_{tt'}:X(t')\to X(t)$, and Eq.(\ref{PullbackP}) now reads (here $x(t)\in X(t), x(t')\in X(t')$):
\begin{eqnarray}
M^*_{t't}(P(t'x(t'))dx^1(t')...dx^D(t')) = \nonumber\\ = P\left(t'M_{t't}(x(t))\right) J(TM_{t't}(x(t)))dx^1(t)...dx^D(t).\label{PullbackP1}
\end{eqnarray}
This shows that the pullback acts in the opposite temporal direction, \begin{eqnarray}
M^*_{t't}: \Omega(X(t')) \to \Omega(X(t)),\label{PullbackExtended}
\end{eqnarray}
as compared to the map inducing it,
\begin{eqnarray}
M_{t't}:X(t)\to X(t').
\end{eqnarray}
Therefore, the pullbacks of inverse maps correspond  to forward propagation in time just as any evolution operator should. This situation can be clarified via the following diagram (here $t>t'>t''$):
\begin{eqnarray}
\begin{array}{ccccc}
t& \stackrel{\text{flow of time}}{\longleftarrow}  & t' & \stackrel{\text{flow of time}}{\longleftarrow}  & t'',\\
X(t) & \stackrel{M_{t't}}{\longrightarrow} &
X(t') & \stackrel{M_{t''t'}}{\longrightarrow} &
X(t''), \\
\Omega(X(t)) & \stackrel{M^*_{t't}}{\longleftarrow} &
\Omega(X(t')) & \stackrel{M^*_{t''t'}}{\longleftarrow} &
\Omega(X(t'')).
\end{array}
\end{eqnarray}
Further, the composition law for pullbacks is,
\begin{eqnarray}
M_{t''t}^* = M^*_{t't} M^*_{t''t'}. \label{CompositionPullback}
\end{eqnarray}
The following equation of infinitesimal temporal evolution holds for pullbacks:
\begin{eqnarray}
\frac\partial{\partial t} M_{t't} = \lim_{\Delta t\to 0} \left(M_{t'(t+\Delta t)}^*-M_{t't}^*\right)/\Delta t= - \hat {\mathcal L}_{\tilde F (t)} M_{t't},\label{EquationLieDerivative}
\end{eqnarray}
where the use has been made of the equality, $M_{t'(t+\Delta t)}^* = M^*_{t(t+\Delta t)} M^*_{t't}$, that follows from Eq.(\ref{CompositionPullback}) and of the definition of the Lie derivative along $\tilde F(t)$:
\begin{eqnarray}
\hat {\mathcal L}_{\tilde F (t)} = \lim_{\Delta t\to0}\left (M^*_{(t+\Delta t)t}-\hat 1_{\Omega(X)}\right)/\Delta t= \tilde F^i(t)\frac\partial{\partial x^i} + \tilde F(t)^i_j dx^j\wedge \hat \imath_{i},
\end{eqnarray}
with $\tilde F^i_j(t) \equiv \partial \tilde F^i(tx)/\partial x^j$, and $\hat \imath_i: \Omega^k(X)\to\Omega^{k-1}(X)$ and $dx^i\wedge: \Omega^k(X)\to\Omega^{k+1}(X)$ being the interior and exterior multiplications respectively. The later can be defined via their actions on the k-form from Eq.(\ref{DefinitionOfForm}) as:
\begin{eqnarray}
dx^i\wedge \psi^{(k)} = (1/k!)\psi_{i_1...i_k} dx^i\wedge dx^{i_1}\wedge...\wedge dx^{i_k},
\end{eqnarray}
and
\begin{eqnarray}
\hat {\imath}_{i} \psi^{(k)} = (1/k!)\sum\nolimits_{j=1}^k  (-1)^{j+1}\psi_{i_1...i_{j-1} i i_{j+1}...i_k}dx^{i_1}\wedge ...\widehat{dx^{i_j}}...\wedge dx^{i_k},
\end{eqnarray}
where the wide hat over the differential denotes that the differential is missing.

Eq.(\ref{EquationLieDerivative}) can be formally integrated with the condition, $M^*_{tt}=\hat 1_{\Omega(X)}$, as:
\begin{eqnarray}
M^*_{t't} = {\mathcal T} e^{-\int_{t'}^{t}d\tau \hat {\mathcal L}_{\tilde F(\tau)}},\label{PullbackIntegrated}
\end{eqnarray}
where $\mathcal T$ is the operator of chronological ordering, which is necessary because $\hat{\mathcal L}_{\tilde F(\tau)}$'s at different $\tau$'s do not commute. Eq.(\ref{PullbackIntegrated}) is a formal notation for the following Taylor series:
\begin{eqnarray}\label{PullbackIntegrated1}
M^*_{t't} = \hat 1_{\Omega(X)} - \int_{t'}^t d\tau_1\hat {\mathcal L}_{\tilde F(\tau_1)} + \int_{t'}^t d\tau_1 \hat {\mathcal L}_{\tilde F(\tau_1)} \int_{t'}^{\tau_1} d\tau_2\hat {\mathcal L}_{\tilde F(\tau_2)} - \dots\label{ChonTaylor}
\end{eqnarray}

\section{Stochastic Dynamics}
\label{StochasticDynamics}

In the previous section, the noise configuration was assumed fixed and the SDE in Eq.(\ref{SDE}) was essentially an ODE so that the dynamics was deterministic. From this point forth, stochastic dynamics will be considered. The goal now is to establish the stochastic (generalized Fokker-Planck) evolution operator known in the dynamical systems theory as the generalized transfer operator.\cite{Rue02} This operator is the evolution operator found in the previous section averaged over the configurations of the noise: 
\begin{eqnarray}
\hat {\mathcal M}_{tt'} = \langle M^*_{t't}\rangle_{Ns}.\label{GTO}
\end{eqnarray}
Note that the stochastic averaging here is legitimate because pullbacks are linear operators on a linear (infinite-dimensional) space, $\Omega(X)$. This operator is always well defined, unlike the concept of stochastically averaged map, which can not be defined if the phase space is not a linear space.   

In Eq.(\ref{GTO}), the following notation for stochastic averaging is introduced:
\begin{eqnarray}
\langle f(\xi) \rangle_\text{Ns} \equiv \iint D\xi f(\xi) P(\xi),\label{DefStochAverage}
\end{eqnarray}
where the functional integration is over all possible noise configurations, $\xi(t)$, and the probability distribution of the noise configurations, $P(\xi)$, is assumed normalized, $\langle 1 \rangle_\text{Ns} = 1$. For simplicity, only Gaussian white noise will be considered,
\begin{eqnarray}
P(\xi) \propto e^{ - \int_{t'}^t d\tau (\xi^a(\tau) \xi^a(\tau) )/2 }, \label{GaussianProbability}
\end{eqnarray}
with the fundamental correlator being,
\begin{eqnarray}
\langle \xi^{a}(\tau_1)\xi^{b}(\tau_2)\rangle_\text{Ns} = \delta^{ab}\delta(\tau_1-\tau_2).\label{GaussianAverage}
\end{eqnarray}

In case of white noise, the noise variables at different times are uncorrelated and $\hat{\mathcal{M}}$ obeys the standard composition law for evolution operators:
\begin{eqnarray}
\hat{\mathcal{M}}_{tt''} = \langle M_{t''t}^* \rangle_\text{Ns} = \langle M^*_{t't} M^*_{t''t'} \rangle_\text{Ns} = \langle M^*_{t't}\rangle_{Ns} \langle M^*_{t''t'} \rangle_\text{Ns} = \hat{\mathcal{M}}_{tt'}\hat{\mathcal{M}}_{t't''},
\label{GTO1}
\end{eqnarray}
where we used the composition law for pullbacks from Eq.(\ref{CompositionPullback}). 

Eq.(\ref{GTO1}) can now be used for the derivation of the Fokker-Planck (FP) equation for the stochastic evolution operator:
\begin{eqnarray}
\partial_t \hat{\mathcal{M}}_{tt'} &=& 
\lim_{\Delta t\to 0} \frac1{\Delta t} \left(\hat{\mathcal{M}}_{(t+\Delta t)t'} - \hat{\mathcal{M}}_{tt'}\right) \nonumber \\&=& 
\lim_{\Delta t\to 0} \frac1{\Delta t} \left(\hat{\mathcal{M}}_{(t+\Delta t)t} - \hat 1_{\Omega(x)}\right) \hat{\mathcal{M}}_{tt'}\nonumber\\&=& - \hat H \hat{\mathcal{M}}_{tt'}.\label{FPEqForM}
\end{eqnarray}
Here, the FP operator is defined as,
\begin{eqnarray}
\hat H = \lim_{\Delta t\to 0} \frac1{\Delta t} \left(\hat 1_{\Omega(x)} - \hat{\mathcal{M}}_{(t+\Delta t)t}\right)=
\lim_{\Delta t\to 0} \frac1{\Delta t} \left(\hat 1_{\Omega(x)} - \langle M^*_{t(t+\Delta t)}\rangle_\text{Ns} \right)\nonumber\\
=
\lim_{\Delta t\to 0} \frac1{\Delta t} \left\langle
\int_t^{t+\Delta t} d\tau_1  \hat{\mathcal{L}}_{\tilde F(\tau_1)} - \int_t^{t+\Delta t} d\tau_1 \hat{\mathcal{L}}_{\tilde F(\tau_1)}\int_t^{\tau_1} d\tau_2 \hat{\mathcal{L}}_{\tilde F(\tau_2)} + ...\right \rangle_\text{Ns},\label{FPOper1}
\end{eqnarray}
where we used Eq.(\ref{ChonTaylor}). Recalling that the Lie derivative is linear in its vector field, 
\begin{eqnarray}
\hat{\mathcal{L}}_{\tilde F(\tau)} = \hat{\mathcal{L}}_{F} + (2\Theta)^{1/2}\xi^a(\tau) \hat{\mathcal{L}}_{e_a},\label{Linearity}
\end{eqnarray}
and using Eqs.(\ref{GaussianAverage}) and $\langle\xi^a(\tau)\rangle_\text{Ns}=0$, one readily arrives at the following expression for the FP operator,
\begin{eqnarray}
\hat H = \hat{\mathcal{L}}_{F} - \Theta \hat{\mathcal{L}}_{e_a}\hat{\mathcal{L}}_{e_a}.\label{FPOp}
\end{eqnarray}
In deriving the above expression for the FP operator, one will find that the second (diffusion) term has the following factor:
\begin{eqnarray}
\lim_{\Delta t\to 0} \frac2{\Delta t} \int_t^{t+\Delta t}d\tau_1 \int_t^{\tau_1}d\tau_2 \delta(\tau_1-\tau_2).
\end{eqnarray}
Here, the upper limit of the integration over $\tau_2$ is right at the "peak" of the $\delta$-function. This integral must be interpreted as $(1/2)$ because no matter how narrow the $\delta$-function is, it is a symmetric function of its argument whereas the integral over the entire domain of its definition is unity. Yet another way of looking at it is to think that the noise is "physical" and its correlation function has a very small, though finite width.     

The integration of Eq.(\ref{FPEqForM}) with the condition, $\hat {\mathcal M}_{tt}=\hat 1_{\Omega(X)}$, gives the following expression for the stochastic evolution operator,
\begin{eqnarray}
\hat {\mathcal M}_{tt'} = e^{-(t-t')\hat H}.
\end{eqnarray}
One can now introduce a time-dependent wavefunction in the representation similar to the Schr\"odinger representation of a quantum theory, $\psi(t)=\hat {\mathcal M}_{tt'}\psi(t')$. The FP Eq.(\ref{FPEqForM}) becomes:
\begin{eqnarray}
\partial_t \psi(t) = - \hat H \psi(t).\label{FPEq}
\end{eqnarray}

\section{Topological Supersymmetry of Stochastic Dynamics}
\label{TopSusy}

One can introduce now the operator of exterior derivative or De Rahm operator:
\begin{eqnarray}
\hat d = dx^i\wedge  \frac\partial{\partial x^i},
\end{eqnarray}
and recall that the Cartan formula expresses the Lie derivative (acting on exterior algebra) as a bi-graded commutator with the exterior derivative, \emph{e.g.}, 
\begin{eqnarray}
\hat {\mathcal L}_G= [\hat d, \hat \imath_G ],\label{CartanFormula}
\end{eqnarray}
where $\hat \imath_G = G^i\hat \imath_i$ is the interior multiplication by some vector field, $G\in TX$, and the bi-graded commutator is defined as:
\begin{eqnarray}
[\hat A, \hat B] = \hat A \hat B - (-1)^{I(\hat A)I(\hat B)} \hat B \hat A,
\end{eqnarray}
with $I(\hat A) = N_A(dx\wedge) - N_A(\hat i)$ being the degree of operator $\hat A$, \emph{i.e.}, the difference between the numbers of exterior and interior multiplication operators in $\hat A$. For example, $I(\hat d)=1$ and $I(\hat \imath_G)=-1$ so that the bi-graded commutator in Eq.(\ref{CartanFormula}) is actually an anticommutator.

The FP operator is $\hat d$-exact, \emph{i.e.}, it has the form of the bi-graded commutator with $\hat d$:
\begin{eqnarray}
\hat H = [\hat d, \hat{\bar d} ],\label{DExactFPOperator}
\end{eqnarray}
where,
\begin{eqnarray}
\hat{\bar d} = F^i\hat{\imath}_i - \Theta e^i_a\hat{\imath}_i\hat {\mathcal L}_{e_a}.\label{dbar}
\end{eqnarray}
In order to establish Eq.(\ref{DExactFPOperator}) one has to make use of the fact that the bi-graded commutator with the exterior derivative is a bi-graded differentiation, \emph{i.e.}, for any two operators $\hat A$ and $\hat B$ the following equality holds,
\begin{eqnarray}
[\hat d, \hat A \hat B] = [\hat d, \hat A] \hat B + (-1)^{I(\hat A)} \hat A [\hat d, \hat B].\label{BiGraded}
\end{eqnarray}
The exterior derivative is bi-graded commutative with any $\hat d$-exact operator. This can be seen from the nilpotency property of the exterior derivative, $\hat d^2=0$, leading to the conclusion that $[\hat d, [\hat d, \hat X]] = 0, \forall \hat X$. Since the FP operator is $\hat d$-exact, one has
\begin{eqnarray}
[\hat d, \hat H] = 0.
\end{eqnarray}
In other words, $\hat d$ is a symmetry of the model. The exterior derivative raises the degree of a wavefunction by one. In the picture where the differentials of a wavefunction are viewed as Grassmann (or fermionic) numbers, this means that $\hat d$ destroys a bosonic variable and substitutes it with a fermionic variable. In other words, $\hat d$ can be recognized as a supersymmetry.

Unconditional existence of this symmetry for all SDEs can be understood as follows: all pullbacks commute with the exterior derivative and so does the stochastic evolution operator, which is nothing else but the stochastically averaged pullback. The physical meaning of this supersymmetry is the preservation of the total probability for the system to exist. Indeed, if $\psi^{(D)}(t')$ is the total probability distribution at time moment $t'$ such that $\int_X \psi^{(D)}(t')= 1$, then at later time moment $t$ the wavefunction is $\psi^{(D)}(t) = \hat{\mathcal M}_{tt'}\psi^{(D)}(t') = \psi^{(D)}(t') + \hat d (\text{ something})$, because $\hat H ... \hat H \psi^{(D)}(t') = \hat d\hat {\bar d}...\hat d\hat {\bar d}\psi^{(D)}(t')$, so that the total probability to exist is unchanged: $\int_X \psi^{(D)}(t)= \int_X (\psi^{(D)}(t') + \hat d(\text{ something}))= 1$. Similar reasoning can be applied to $\hat d$-closed wavefunctions of lesser degree, which can be interpreted as the conditional probability distributions of variables that are statistically independent from the variables in which the wavefunction is not a differential form.

The most general class of models that possess this type of supersymmetry is the Witten-type topological or cohomological theories (ChT). \cite{Bir91,Lab89,Wit88,Wit881,Fre07} The STS can not be identified, however, as a full-fledged ChT because in a typical ChT one limits his interest only to the supersymmetric states. In the STS, one is interested primarily in the global ground states that are not supersymmetric for chaotic models.  

From the group-theoretic point of view, the topological supersymmetry is a continuous one-parameter group of transformations on $\Omega(X)$,
\begin{eqnarray}
\hat G_\alpha = (\hat G_{-\alpha})^{-1} = e^{\alpha\hat d} = 1 + \alpha \hat d, \alpha \in \mathbb{R}^1,
\end{eqnarray}
and the FP operator is invariant with respect to these transformations,
\begin{eqnarray}
\hat G_\alpha \hat H \hat G_{-\alpha} = \hat H.
\end{eqnarray}
Just as in case of any other symmetry, the eigenstates of $\hat H$ must be irreducible representations of this group. There are only two types of irreducible representations of $\hat d$: most of the eigenstates are non-$\hat d$-symmetric "bosonic-fermionic" doublets or pairs of eigenstates, whereas some of the eigenstates are $\hat d$-symmetric singlets. The eigensystem of $\hat H$ will be discussed in more details in the next section. At this point, it is worth mentioning that not all $\hat d$-closed operators, \emph{i.e.}, operators that commute with $\hat d$, are necessarily $\hat d$-exact as in Eq.(\ref{DExactFPOperator}). In fact, a $\hat d$-exact evolution operator implies not only commutativity with $\hat d$, but also that all the $\hat d$-symmetric eigenstates of $\hat H$ have zero eigenvalues.

\section{Fokker-Planck Spectrum}
\label{Spectrum}

Our primary goal in this section is to discuss the eigensystem of the FP operator. Before we get to this task, allow us to discuss the FP operator from a more general perspective. 

No approximations were made in the derivation of the FP operator in Sec.\ref{StochasticDynamics}. This is the correct and unique evolution operator that describes stochastic evolution of differential forms in accordance with Eq.(\ref{SDE}). This resolves various ambiguities about the FP operator that can be found in the literature. One of such ambiguities is known as the Ito-Stratonovich dilemma.\cite{VanKampen,Moo14} Its essence is that the Ito and Stratonovich stochastic calculii lead to different evolution operators. \footnote{In Ref.\cite{Ovc13}, is was shown that the Ito-Stratonovich dilemma is related to the operator ordering ambiguity that appears every time one passes from the pathintegral representation of quantum or stochastic dynamics to its operator representation. At this, the correct Stratonovich approach corresponds to the bi-graded Weyl symmetrization rule.} The correct form of the FP operator for the total probability distribution is obtained from Eq.(\ref{FPOp}) by noting that the Lie derivative acts on top differential forms as, $\hat{\mathcal L}_{G} \psi^{(D)}(x) = \partial/\partial x^i G^i(x) \psi^{(D)}(x)$. Thus:
\begin{eqnarray}
\partial_t \psi^{(D)} = - \left(\frac\partial{\partial x^i} F^i- \Theta \frac\partial{\partial x^i} e^i_a\frac\partial{\partial x^j} e^j_a \right)\psi^{(D)}.\label{StratonovichFP}
\end{eqnarray}
This is the well-known Stratonovich FP equation for the total probability distribution. 

One possible confusion about the stochastic evolution operator is as follows. The point is that the stochastic quantization is a close relative with the supersymmetric nonlinear sigma models or Witten models. \cite{Wit82} In the later class of models, the kinetic term (or rather the diffusive term in case of stochastic dynamics) of the evolution operator is the Hodge Laplacian. Therefore, it may seem reasonable to believe that the evolution operator of the STS is:
\begin{eqnarray}
\hat H_{Hodge} = \hat{\mathcal L}_F + \Theta [\hat d, \hat d^\dagger].
\end{eqnarray}
where the adjoint of the exterior derivative is $\hat d^\dagger=-g^{ij} \hat \imath_{i} \left( \partial_j + g_{kp} (g^{pm})_{'j} dx^k \wedge \hat \imath_{m} + (1/2) log(g)_{'j} \right)$, with $g=\det g_{ij}$ and $ _{'j}$ denoting differentiation. For general $e$'s, however, this is not so,
\begin{eqnarray}
\hat H \ne \hat H_{Hodge}.
\end{eqnarray}
The diffusion term is a member of the family of Laplace operators,
\begin{eqnarray}
-\hat{\mathcal{L}}_{e_a}\hat{\mathcal{L}}_{e_a} = g^{ij}(x)\frac\partial{\partial x^i}\frac\partial{\partial x^j} + ... \label{DiffLapl}
\end{eqnarray}
However, unless one considers, \emph{e.g.}, the position-independent $e$'s, Eq.(\ref{DiffLapl}) is not the Hodge Laplacian, $[\hat d, \hat d^\dagger]$. \footnote{Neither it is the Bochner (or Bertlami) Laplacian, $-\hat \nabla_i^\dagger g^{ij}\hat \nabla_j$ (with $\hat \nabla_i$ being the covariant derivative), which is also sometimes mistakenly believed to be the diffusion part of the stochastic evolution operator.} Nevertheless, just as the Hodge Laplacian, the diffusion Laplacian of the FP operator possesses the important property of being $\hat d$-exact, $[\hat d, -e^i_a\hat\imath_i\hat{\mathcal L}_{e_a}]$.

Let us turn to the discussion of the eigensystem of the FP operator. For simplicity, it is assumed that the phase space is closed, the noise-induced metric $g^{ij}=e^i_ae^j_a$ is positive definite everywhere, and the temperature $\Theta>0$ so that the FP operator is elliptic. We find it it reasonable to believe, however, that most of the claims in this section hold true or at least transformative to more general classes of models.

First of all, the FP operator is real and consequently its spectrum consists of real eigenvalues and pairs of complex conjugate eigenvalues that in the DS theory are known as the Ruelle-Pollicott resonances (RP resonances). This means that $\hat H$ is pseudo-Hermitian. \cite{Mos02} Therefore, the eigensystem constitutes a complete bi-orthogonal basis in the Hilbert space,
\begin{subequations}
\label{Eigensystem}
\begin{eqnarray}
\hat H |\psi_n\rangle = \mathcal{E}_n |\psi_n\rangle, && \langle \psi_n| \hat H  = \langle \psi_n| \mathcal{E}_n , \\
\sum\nolimits_n |\psi_n\rangle \langle \psi_n | = \hat 1_{\Omega(X)}, &&\langle \psi_n |\psi_m \rangle  =  \int_X \psi_m \wedge \bar \psi_n =  \delta_{nm}.
\end{eqnarray}
\end{subequations}
Here we introduced the bra-ket notation: $|\psi_n \rangle \equiv \psi_n$ and $\langle \psi_n | \equiv \bar \psi_n$. The integration in the last formula over $X$ is nonzero only if $\psi_k \wedge \bar \psi_n\in\Omega^D(X)$. This suggests that the bra-ket combination of any eigenstate is a top differential form, $\psi_n \wedge \bar \psi_n \in \Omega^D(X)$, having the meaning of the total probability distribution associated with this eigenstate. Because the FP operator is non Hermitian, the relation between bra's and ket's is not trivial: $\bar \psi_n = \sum_{m} \star(\psi_m^*) \eta_{mn}$, where $\eta_{mn}$ is the metric on the Hilbert space and $\star$ denotes Hodge conjugation.

As already mentioned in the previous section, all the eigenstates are divided into two groups: the pairs of non-$\hat d$-symmetric eigenstates and the singlets of $\hat d$-symmetric eigenstates. Each pair of non-$\hat d$-symmetric eigenstates can be defined via a single bra-ket pair, $|\tilde\vartheta_n\rangle$ and $\langle\tilde\vartheta_n|$, such that $\langle\tilde\vartheta_n|\hat d |\tilde \vartheta_n \rangle = 1$. The bra-ket pairs of the non-$\hat d$-symmetric pairs of eigenstates are given as,
\begin{eqnarray}
&|\vartheta_n \rangle = |\tilde\vartheta_n\rangle, \text{ } \langle\vartheta_n| = \langle\tilde\vartheta_n|\hat d, \label{FirstType}\\
&\text{and},\nonumber\\
&|\vartheta'_n \rangle = \hat d |\tilde\vartheta_n\rangle, \text{ } \langle\vartheta'_n| = \langle\tilde\vartheta_n|.\label{SecondType}
\end{eqnarray}
The $\hat d$-symmetric singlets, that can be called $|\theta\rangle$'s, are such that
\begin{eqnarray}
\hat d|\theta_k\rangle=0, \text{ but }|\theta_k\rangle \ne \hat d|\text{something}\rangle,\label{Singlets}
\end{eqnarray}
and the same for the bra, $\langle\theta_k|\hat d=0$, but $\langle\theta_k| \ne \langle \text{something}|\hat d$.
Eq.(\ref{Singlets}) is nothing else but the condition for a state to be non-trivial in De Rahm cohomology.

All expectation values of $\hat d$-exact operators vanish on $\hat d$-symmetric states:
\begin{eqnarray}
\langle\theta_k| [\hat d, \hat X] |\theta_l\rangle = 0.\label{ZeroExactOperators}
\end{eqnarray}
The FP operator is also $\hat d$-exact so that all $\hat d$-symmetric eigenstates have zero-eigenvalues, $0=\langle\theta_k| \hat H|\theta_k\rangle = \mathcal{E}_{\theta_k}\langle\theta_k| \theta_k\rangle = \mathcal{E}_{\theta_k}$.

Each De Rahm cohomology class must provide one $\hat d$-symmetric eigenstate. This can be seen in the deterministic limit. In this limit, the FP operator is the Lie derivative along the flow vector field and the ket's of the supersymmetric states are Poincar\'e duals\footnote{
A Poincar\'e dual of a submanifold is a $\delta$-function like distribution of this submanifold with differentials only in the transverse directions.} of the global unstable manifolds.\footnote{The bra's of these states are Poincar\'e duals of the global stable manifolds. The bra's and ket's of these eigenstates intersect on invariant manifold. In case when invariant manifolds are not points, as in Morse-Bott case, the eigenstates must be complemented by elements from the De Rahm cohomology of the invariant manifolds.} For integrable or non-chaotic flow vector fields, the global unstable manifolds are well-defined topological manifolds. Each homology class of $X$ may contain more than one global unstable manifold with the corresponding supersymmetric eigenstate from the same De Rahm cohomology class. At this, the multiple eigenstates from the same De Rahm cohomology class will differ by $\hat d$-exact pieces. This situation happened because in the deterministic limit some of the non-$\hat d$-symmetric eigenstates accidentally have zero-eigenvalues. 

If one introduces a weak noise, the wavefunctions of these eigenstates will get smeared in the transverse directions thus introducing the tunneling matrix elements between the wavefunctions on different "parallel" global unstable manifolds. This tunneling will remove the above accidental degeneracy leaving only one supersymmetric eigenstates in each De Rahm cohomology class. If it is not so, the eigensystem of $\hat H$ is incomplete, whereas the eigensystem of a pseudo-Hermitian elliptic operator must be complete on a compact $X$.

Whether or not each De Rahm cohomology class provides a supersymettric eigenstate is not that important for our discussion. What is important is the unconditional existence of at least one $\hat d$-symmetric eigenstate. This eigenstate is from $\Omega^{D}$. Its existence can be established through the physical version of the completeness argument. Indeed, all the non-$\hat d$-symmetric eigenstates from $\Omega^D(X)$ are $\hat d$-exact, \emph{i.e.}, of type (\ref{SecondType}). The integral of all such non-$\hat d$-symmetric eigenstates over $X$ is zero, $\int_{X} \hat d \tilde\vartheta = 0$. On the other hand, a wavefunction from $\Omega^D(X)$ has the meaning of the total probability distribution. The integral of a physically meaningful total probability distribution over $X$ must not vanish. This implies that at least one $\hat d$-symmetric eigenstate from $\Omega^D(X)$ must exist if the model is physically meaningful. This supersymmetric eigenstate must be recognized as the steady-state (zero eigenvalue) total probability distribution of the thermodynamic equilibrium of the model.

\begin{figure}
\centerline{\includegraphics[height=3.3cm, width=10cm]{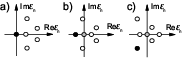}}
\caption{\label{Figure2} Possible spectra of the FP operator. $\hat d$-symmetry is spontaneously broken for {\bf b} and {\bf c}, because the ground states (black dots) have non-zero eigenvalues and thus are non-$\hat d$-symmetric. The pseudo-time reversal symmetry is spontaneously broken in {\bf c}, because the ground state is one out of two Rulle-Pollicott resonances with the least attenuation rate. Dots with grey filling at the origin in {\bf b} and {\bf c} represent $\hat d$-symmetric states that are not ground states of the model anymore.}
\end{figure}

The above properties of the FP spectrum limit the possible FP spectra to only three types presented in Fig.\ref{Figure2}.
In the DS theory, there are theorems (see, \emph{e.g.}, proposition 3 of Ref.\cite{Rue02} about the spectra of the generalized transfer operator in Eq.(\ref{GTO})) stating that in certain classes of models the eigenstate with the least attenuation rate is real. There are no theorems, however, that would claim that spectra of type given in Fig.\ref{Figure2}c are not realizable in principle. Therefore, such models must also be analyzed.

Models with spectra of types given in Figs.\ref{Figure2}b and \ref{Figure2}c have their topological supersymmetry spontaneously broken because the ground states have non-zero eigenvalues and thus are non-$\hat d$-symmetric. Furthermore, as a pseudo-Hermitian operator, the FP operator also possesses the pseudo-time-reversal symmetry (or $\eta T$-symmetry),with respect to which the RP resonances are the $\eta T$-companions.\cite{Mos02} For the spectra in Fig.\ref{Figure2}c, both RP resonances with the lowest attenuation rate are candidates for the title of the ground state of the model. In the next section it will be discussed that in order to preserve the ergodicity property, \emph{i.e.}, the property of the reduction in the long time limit of the stochastic average to the average over the ground states only, one must choose one of these RP resonances as the ground state. This will spontaneously break the $\eta$T-symmetry. This phenomenon, if it is physically realizable, may be interesting in its own right. We will discuss it here any further, however.

\section{Dynamical Partition Function and Witten Index}
\label{SecPartitionFunction}

One of the most fundamental objects in the theory under consideration is the dynamical partition function (DPF), 
\begin{eqnarray}
Z_{t0} = Tr e^{-t\hat H}. \label{PartitionFunction}
\end{eqnarray}
This object is known in the dynamical systems theory as the counting trace of the generalized transfer operator. \cite{Rue02} It can be shown that for a certain class of models and in the limit of the infinitely long temporal propagation, the DPF represents the stochastically averaged number of periodic solutions of the SDE. \cite{Ovc14} 

Let us recall now that in deterministic chaotic models the number of periodic solutions grows exponentially with time in the large time limit. The contribution comes from infinite number of unstable periodic orbits with arbitrary large periods that constitute "strange" attractors. \cite{Gil98} This is the reason why chaotic dynamics are also called sometimes "complex" dynamics. This term is borrowed from information theory, where a problem is viewed "complex" if the number of elementary operations needed for its solution grows exponentially with the "size" of the problem.

Stochastic generalization of this situation is given in Fig.\ref{Figure2}b. There, in the limit of infinitely long temporal evolution, only the ground state with the lowest attenuation rate contribute to the DPF, which thus grows exponentially,
\begin{eqnarray}
\left.Z_{t0}\right|_{t\to \pm\infty} \approx 2 e^{|{\mathcal E}_g|t},\label{ExponentialGrowth}
\end{eqnarray}
with ${\mathcal E}_g$ being the real (and negative) eigenvalue of the ground state(s) and the factor $2$ comes from the two-fold $\hat d$-degeneracy of this state. The ground state is non-$\hat d$-symmetric for it has non-zero eigenvalue. Thus, the topological supersymmetry is broken spontaneously in this case. This is the first indication on that the stochastic generalization of dynamical chaos is the phenomenon of spontaneous breakdown of topological supersymmetry. The emergence of the long-range response (the BF effect) in Sec.\ref{SecButterflyEffect} and the brief discussion of the structure of the ground states in the deterministic limit in Sec.\ref{States} below are two additional indication that this identification is indeed correct.  

It is also important to discuss the fundamental difference between the \emph{dynamical} partition function in Eq.(\ref{PartitionFunction}) and the \emph{thermodynamical} partition function of statistical quantum physics. The latter is defined as $Tr e^{- \beta \hat{H}_q }$, where $\hat H_q$ is a Hermitian Hamiltonian of a quantum model under consideration and $\beta$ is the inverse temperature. This thermodynamical partition function and Eq.(\ref{PartitionFunction}) have very similar appearances. Moreover, in the Literature on, \emph{e.g.}, N=2 supersymmetric quantum mechanics, it is often said that the DPF (\ref{PartitionFunction}) is the result of the Wick rotation of the "real" time of temporal evolution, which can be symbolically expressed as $t\to i\times t \sim \beta$. This is not quite correct from the point of view of stochastic dynamics. The time in Eq.(\ref{PartitionFunction}) is the original time of the temporal evolution in the SDE (\ref{SDE}) and/or the stochastic FP evolution. The direct quantum analogue of Eq. (\ref{PartitionFunction}) is the trace of the quantum evolution operator or the generating functional, $Tr\hat U_{t0} = Tr e^{-i t \hat H_q}$.

The fact that for a certain class of SDEs the DPF represents in the long time limit the stochastic number of periodic solutions of the SDE has yet another important implication. It suggests that the entire exterior algebra must be viewed as the Hilbert space and not only of the total probability distributions as is classical approaches to SDEs. Without this assumption, the DPF of any SDE does not represent the stochastically averaged number of periodic solutions, which, in its turn, is the closest physical meaning for the DPF.

Yet better proof that the entire exterior algebra must be viewed as the Hilbert space of SDEs is the physical meaning of the Witten index known in the dynamical systems theory as the sharp trace of the generalized transfer operator.\cite{Rue02}
\begin{eqnarray}
W_{t0} \equiv W = Tr (-1)^{\hat F}e^{-t\hat H},
\end{eqnarray} 
where $\hat F = dx^i \wedge \hat \imath_i$ is the operator of the degree of the differential form and/or the wavefunction. $\hat F$ is commutative with the stochastic evolution operator so that its eigenvalue is a "good" quantum number. This suggests in particular that all the non-$\hat d$-symmetric eigenstates, which come in pairs with degree differing by one, do not contribute to $W$. Only the $\hat d$-symmetric eigenstates, which all have zero eigenvalues, contribute to the Witten index. Thus, $W$ is independent of the time of stochastic evolution as well as of many other things. This quantity is, in fact, of topological origin. Its mathematical meaning is the stochastically averaged Lefschetz index of the maps defined by the SDE.\cite{Ovc14} Up to a topological constant, it represents the partition function of the noise.\cite{Ovc13} Clearly, if one would not be thinking of the entire exterior algebra of $X$ as of the Hilbert space of the stochastic model, such a fundamental object (the partition function of the noise) would not have a representative within the theory under consideration.  

The FP spectrum given in Fig.\ref{Figure2}c does not comply with the idea that the DPF represents the stochastically averaged number of periodic solutions of the SDE. Indeed, in the large time limit, the DPF of such a model is, 
\begin{eqnarray}
\left.Z_{t0}\right|_{t\to+\infty} \approx 4 \cos ( \text{Im}{\mathcal E}_g t) e^{|\text{Re}{\mathcal E}_g| t },
\end{eqnarray}
where the contribution comes from the two RP resonances (each doubly degenerate) with the lowest attenuation rate. It takes on negative values and a negative number of periodic solutions makes no sense. This does not necessarily imply, however, that such spectra are not realizable. It only suggests that in this particular situation the DPF can not be interpreted as the stochastically averaged number of periodic solutions. In the next section, this situation will be discussed in more details. 

\section{Response and the Butterfly Effect}
\label{SecButterflyEffect}

There is a special class of correlators that represent the response of the model to external perturbations. It is understood that the only "physical" way to couple a model to an external influence is on the level of the SDE itself. The coupling can be realized by the modification of the flow vector field with the external probing fields, $\phi^c(t), c=1,2...$, in the following manner (see Eq.(\ref{SDE})):
\begin{eqnarray}
\tilde F(t) \to \tilde F(t) - \phi^c(t) f_c,
\end{eqnarray}
where $f_c(x)\in TX_x, c = 1, 2, ...$ is a set of vector fields on $X$. The stochastic evolution operator will change as:
\begin{eqnarray}
\hat{\mathcal M}_{tt'} = \langle {\mathcal T}e^{-\int_{t'}^{t}d\tau\left(\hat{\mathcal L}_{\tilde F(\tau)} - \phi^c(\tau) \hat{\mathcal L}_{f_c(\tau)}\right)}\rangle_{Ns}.
\end{eqnarray}
One can now introduce the generating functional:
\begin{eqnarray}
Z_{tt'}(\phi) = Tr \langle {\mathcal T} e^{-\int_{t'}^{t}d\tau\left(\hat{\mathcal L}_{\tilde F(\tau)} - \phi^c(\tau) \hat{\mathcal L}_{f_c(\tau)}\right)} \rangle,
\end{eqnarray}
with the help of which the response of the model to external influence is characterized by the following stochastic expectation values and/or correlators ($t_k>t_{k-1}...>t_1$):
\begin{eqnarray}
R_{c_k...c_1}(t_k...t_1) &=& \lim_{t_\pm \to \pm\infty} \left. Z_{t_+t_-}^{-1} \frac{\delta^k }{\delta \phi_{c_k}(t_k)...\delta \phi_{c_1}(t_1)} Z_{t_+t_-}(\phi)\right|_{\phi=0}\nonumber\\
&=&\lim_{t_\pm \to \pm\infty} Z_{t_+t_-}^{-1} \langle e^{-\int_{t_k}^{t_+}d\tau \hat{\mathcal L}_{\tilde F(\tau)} } \hat {\mathcal L}_{c_k}e^{-\int_{t_{k-1}}^{t_k}d\tau \hat{\mathcal L}_{\tilde F(\tau)} } ... \nonumber\\&& ... e^{-\int_{t_1}^{t_2}d\tau \hat{\mathcal L}_{\tilde F(\tau)} }\hat {\mathcal L}_{f_{c_1}} e^{-\int_{t_-}^{t_1}d\tau \hat{\mathcal L}_{\tilde F(\tau)} }\rangle_{Ns}.\nonumber
\end{eqnarray}
For models with spectra of types in Figs.\ref{Figure2}a and \ref{Figure2}b, one can perform now the stochastic averaging arriving at:
\begin{eqnarray}
R_{c_k...c_1}(t_k...t_1) &=& \lim_{t_\pm \to \pm \infty}Z_{t_+t_-}^{-1}\sum_{n} \langle n| \hat {\mathcal M}_{t_+ t_k} \hat {\mathcal L}_{c_k}\hat {\mathcal M}_{t_{k} t_{k-1}}  ... \hat {\mathcal M}_{t_2 t_1} \hat {\mathcal L}_{c_1}\hat {\mathcal M}_{t_1 t_-}| n\rangle\nonumber\\
&=& N_g^{-1} \sum_{g}\langle g| \hat {\mathcal M}_{t_- t_k} \hat {\mathcal L}_{c_k}\hat {\mathcal M}_{t_{k} t_{k-1}}  ... \hat {\mathcal M}_{t_2 t_1} \hat {\mathcal L}_{c_1}\hat {\mathcal M}_{t_1 t_-} | g\rangle. \label{ergodicity}
\end{eqnarray}
Here, $g$ runs over the ground states with the least attenuation rate, $N_g$ is the number of the ground states, and the following equality has been used
\begin{eqnarray}
\lim_{t_\pm \to \pm \infty} Z_{t_+t_-}^{-1} \sum_{n} \langle n| \hat {\mathcal M}_{t_+ t_k} = \lim_{t_- \to - \infty} N_g^{-1} \sum_{g} \langle g| \hat {\mathcal M}_{t_- t_k},\label{InterEq}
\end{eqnarray}
which follows from the observation that all the other non-ground eigenstates will provide only exponentially vanishing contribution, as well as from the following identities, 
\begin{eqnarray}
\left.Z_{t_+t_-}\right|_{t_\pm \to \pm \infty} = N_g e^{-{\mathcal E}_g (t_+-t_-)},
\end{eqnarray}
and
\begin{eqnarray}\nonumber
\left. Z_{t_+t_-}^{-1} \langle g| \hat {\mathcal M}_{t_+ t_k}\right|_{t_\pm \to \pm \infty}  = \left. \langle g| e^{-{\mathcal E}_g (t_k-t_-)}\right|_{t_- \to = -\infty} = \left.\langle g| \hat {\mathcal M}_{t_- t_k}\right|_{t_- \to -\infty}.
\end{eqnarray}

The equality (\ref{ergodicity}) can be called the ergodicity property\footnote{Here we use a generalized version of the concept of ergodicity. In the Literature, ergodicity is often equivalent to what in this paper corresponds to thermodynamic equilibrium and/or the unbroken topological supersymmetry.} of models with spectra in Figs.\ref{Figure2}a and \ref{Figure2}b. This is the property that the stochastic expectation values/correlators automatically reduce in the limit of infinitely long temporal evolution down to the expectation values/correlators over the ground states only.\footnote{In quantum theory, such expectation values are known as vacuum expectation values} 


\begin{figure}
\centerline{\includegraphics[height=2.65cm, width=12cm]{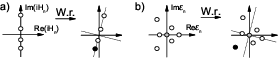}}
\caption{\label{Figure3} {\bf (a)} In quantum theory, the ground state is the one with the lowest energy. This can be justified by Wick rotating time a "little", $t \to t + i0 $, so that the trace of the evolution operator, $\left.Z_t\right|_{t\to\infty} = Tr e^{-i\hat H_q t} \to e^{-i E_g t}$, where $E_g$ is the lowest or the ground state eigenvalue. This is one of the ways to validate the concept of the vacuum expectation value in quantum theory.  {\bf (b)} Just as convincing should sound the idea that in the STS of models where a pair of RP resonances has the least attenuation rate one can declare one of these states to be the ground state. In result, the stochastic expectation values become the ground state expectation value.}
\end{figure}

For models with spectra in Fig.\ref{Figure2}c, the ergodicity property in Eq.(\ref{ergodicity}) is not "automatic". Just like in quantum theory, there is a need for an additional argument in order to transform the expectation value into the expectation value over the ground state(s) only. One of such arguments is given in Fig.\ref{Figure3}: one can Wick rotate time a "little" so that Eq.(\ref{ergodicity}) becomes correct. It can be said that the use of this procedure of choosing one out of two PR resonances as the ground state of the model is the phenomenon of the spontaneous breakdown of the $\eta$T symmetry (see the end of Sec.\ref{Spectrum}) because the two RP resonances with the lowest attenuation rate are $\eta$T-companions and we choose only one of them over the other.

In analogy with the quantum theory, one can introduce now the Heisenberg representation of operators,
\begin{eqnarray}
\hat A (t) = \hat {\mathcal M}_{t_- t} \hat A \hat {\mathcal M}_{t t_-}. \label{Heisenb} 
\end{eqnarray}
In this picture, one has,
\begin{eqnarray}
R_{c_k...c_1}(t_k...t_1) &=& N_g^{-1}\sum_g\langle g| {\mathcal T} \hat {\mathcal L}_{c_k}(t_k) ... \hat {\mathcal L}_{c_1}(t_1)| g\rangle,\label{Expect1}
\end{eqnarray}
where the previous condition, $t_k>...>t_1$, is relaxed and the chronological ordering operator is introduced instead. Note that technically Eq.(\ref{Heisenb}) is dependent on $t_-$, whereas Eq.(\ref{Expect1}) does not. In fact, because of the time-translation invariance of the theory, Eq.(\ref{Heisenb}) is dependent only on time differences of its arguments and, in particular, all
\begin{eqnarray}
R_{c_1}(t_1) &=& N_g^{-1}\sum_g\langle g| \hat {\mathcal L}_{c_1}| g \rangle, \label{Expect2}
\end{eqnarray}
are independent of time.

Using now that the exterior derivative is commutative with the stochastic evolution operator so that
\begin{eqnarray}
\hat d(t) \equiv \hat d,
\end{eqnarray}
together with the Cartan formula (\ref{CartanFormula}) and the fact that the bi-graded commutator with $\hat d$ is a bi-graded differentiation, one has:
\begin{eqnarray}
R_{c_k...c_1}(t_k...t_1) = N_g^{-1}\sum_g\langle g|\left[\hat d, \hat R_{c_k...c_1}(t_k...t_1)\right]  |g\rangle,\label{Response}
\end{eqnarray}
where
\begin{eqnarray}
\hat R_{c_k...c_1}(t_k...t_1) = {\mathcal T} \left(\hat \imath_{f_{c_k}}(t_k) \hat {\mathcal L}_{c_{k-1}}(t_{k-1})... \hat {\mathcal L}_{c_1}(t_1)\right).
\end{eqnarray}

When the ground states are $\hat d$-symmetric, \emph{i.e.}, the topological supersymmetry is not broken and the model can be said to be at thermodynamic equilibrium, the response correlators in Eq.(\ref{Response}) vanish by the definition of the $\hat d$-symmetric states in Eq.(\ref{ZeroExactOperators}). In other words, in the long-time limit, the model does not respond to or rather "forgets" the perturbations. On the contrary, if the topological supersymmetry is spontaneously broken and the ground states are non-$\hat d$-symmetric, some of the response correlators do not vanish. This means that the model remembers perturbations even in the limit of the infinite duration of temporal evolution. This is how the STS reveals the emergence of the famous butterfly effect in chaotic SDEs.

\section{Deterministic Ground States}
\label{States}
The emergence of the BF effect in models with spontaneously broken topological supersymmetry is one of the indications on that the chaotic behavior must be identified with the spontaneous breakdown of topological supersymmetry. In this section, this claim is also supported by the brief analysis of the structure of the ground states in the deterministic limit.

\begin{figure}[t!]
\centerline{\includegraphics[height=6cm, width=9cm]{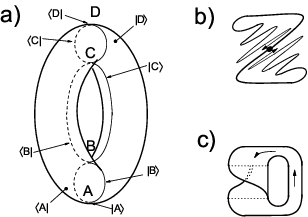}}
\caption{\label{Fig4} The global $\hat d$-symmetric ground states of a Langevin SDE on a torus in the deterministic limit and with the Langevin potential being the "height". Each ground state is the Poincar\'e dual of a global unstable manifold of one of the four critical points (A,B,C,D). The bra's of the ground state are the Poincar\'e duals of the corresponting stable manifolds. The bra-ket combination of each ground state is a delta-functional distribution on the four critical points. {\bf (a)} The qualitative representation of (the Poincar\'e section) of the unstable manifold of the deterministic chaotic behavior known as homoclinic tangle. The unstable manifold folds on itself in a recursive manner and this folding accumulates at the origin. There, the orientation of the unstable manifold is not well-defined and the wavefunction must vanish as is indicated by the fading width of the curve. This suggests that such a ground state is non-$\hat d$-symmetric and that the topological supersymmetry is spontaneously broken. {\bf (b)} The effective representation of the unstable manifold of the Roessler model in the topological theory of chaos (see, e.g., Ref.\cite{Gil98}). The unstable manifold is a branched manifold that has self-intersection. An attempt to construct a Poincar\'e dual will lead to either the requirement that the wavefunction vanishes at the self-intersection or to the effective existence of a boundary at the self-intersection. In either case, the corresponding wavefunction is non-$\hat d$-symmetric.}
\end{figure}
%

In the deterministic limit, all models are divided into non-chaotic and chaotic ones depending on whether or not the flow vector field is integrable or non-integrable in the sense of the dynamical systems theory.\footnote{The flow vector field is called integrable in the sense of the DS theory if its global unstable (and stable) manifolds are well-defined foliations of the phase space.} For models with integrable flows, the ket's of the supersymmetric ground states are the so-called Poincar\'e duals of the global unstable manifolds. The Poincar\'e duals are the $\delta$-functional distributions on the unstable manifolds with the differentials in the transverse directions. For example, the Poincar\'e duals of two lines given on the $x-y$ plane as $x=const$ and $y=const$ are respectively, $\delta(x-const)dx$ and $\delta(y-const)dy$. 

It is straightforward to show that Poincar\'e duals of global unstable manifolds are unchanged by the flow, \emph{i.e.}, they are zero-eigenvalue eigenfunctions of the FP operator, which in the deterministic limit is the Lie derivative along the flow. Furthermore, it is also easy to see that since the global unstable manifolds do not have boundaries, these wavefunctions are $\hat d$-symmetric. The textbook example of the global supersymmetric eigenstates for Langevin SDE on torus in the deterministic limit is given in Fig.\ref{Fig4}a.

Each homology class of the phase space may contain more than one global unstable manifold and each of these manifolds corresponds to a $\hat d$-symmetric ground state. Each of these ground states is a superposition of one $\hat d$-symmetric ground state non-trivial in the corresponding De Rahm cohomology class and a $\hat d$-exact piece. This means that pairs of non-$\hat d$-symmetric eigenstates accidentally have zero eigenvalues. This accidental degeneracy will be lifted by any weak noise, which will introduce exponentially weak tunneling processes between the states (instanton-antiinstanton configuration). In result, at non-zero noise, each De Rahm cohomology will have only one $\hat d$-symmetric ground state.

Let us now turn to chaotic or non-integrable deterministic models. We recall that in integrable case, invariant manifolds lay on the intersection of stable and unstable manifolds. Similar situation occurs in chaotic deterministic models. There, strange or chaotic attractors are formed by the intersection of the stable and unstable manifolds and the bra/ket of the ground state's wavefunction must represent (or rather be) the Poincar\'e duals of these manifolds.

In chaotic models, however, (un)stable manifolds are not well defined topological manifolds. They can fold on themselves in a recursive manner as qualitatively illustrated for "homoclinic tangle" in Fig.\ref{Fig4}b. As a result, the straightforward attempt to come up with a Poincar\'e dual for such an unstable manifold leads to the ambiguity in the orientation of the manifold at the point of the accumulation of self-folding.

This situation has its analogues in quantum theory. For example, non rotationally symmetric electron wavefunction on a rotationally symmetric atom (p-,d-,f-... orbitals) will have an ambiguity at the origin unless it vanishes there. Another example is from the theory of type II superconductors, where the superconducting order parameter at the core of an Abrikosov vortex must vanish because otherwise its value will be ambiguous. In our case, the ambiguity of the Poincar\'e dual of the unstable manifold of a homoclinic tangle can be remedied by modifying it with a continuous function that goes to zero at the "origin". At this, the coordinate dependence along the manifold would automatically suggest that the wavefunction is not annihilated by $\hat d$. In other words, the wavefunction of the ground state is not $\hat d$-symmetric.

Another way to see that in chaotic deterministic models the ground state is not $\hat d$-symmetric is provided by the topological theory of chaos. \cite{Gil98} In this theory, the unstable manifold is qualitatively approximated by a branched manifold that has self-intersections as illustrated for the case of R\"ossler model in Fig.\ref{Fig4}c. Acting by $\hat d$ on the formal Poincar\'e dual of this branched manifold gives the Poincar\'e dual of its self-intersection. That is, such wavefunction is not $\hat d$-symmetric either. 

In this manner, the above simple analysis of the structure of the ground states in the deterministic limit also points out onto that the dynamical chaos is indeed the phenomenon of the spontaneous topological supersymmetry breaking. In combination with the finding of the emergence of the long-range response/memory in models with spontaneously broken supersymmetry in the previous section \ref{SecButterflyEffect}, it provides ample evidence that the concept of dynamical (stochastic) chaos is indeed the phenomenon of the spontaneous breakdown of topological supersymmetry. 

\section{Conclusion}
\label{SecConclusion}

In this paper, it is shown that the approximation-free picture of chaotic dynamics as a topological supersymmetry breaking explains the emergence of the butterfly effect and generalizes the concept of dynamical chaos to stochastic dynamics. These two results were not available in the previous theoretical approaches.

Surprisingly enough, a symmetry breaking picture of (stochastic) chaotic dynamics is somewhat contradictory with the semantics of the word "chaos", which in the daily language means the "absence of order". Within the STS, it is actually the chaotic phase which is the low-symmetry or "ordered" phase. Unlike models with unbroken topological supersymmetry (thermodynamic equilibrium), chaotic models possess a long-range order that reveals itself not only though the BE effect as is discussed in this paper but also through such well-established long-range natural dynamical phenomena as the 1/f noise and the power-law statistics (flicker noise) of various avalanches-type processes - the very reason why the Richter scale for the earthquakes' magnitudes is logarithmic.

As many other symmetry breaking phenomena, stochastic chaos must admit the description in terms of the low-energy effective theory (LEET) for the order parameter, \emph{i.e.}, the collection of only "important" variables. For chaos, these important variables are the unstable (or unthermalized) variables of the ground state in which the system has infinite memory of perturbations and the LEET must be written for the gapless fermions (goldstinos) that are the supersymmetric partners of the unstable variables. We find it a rather interesting and nontrivial finding that the appropriate low-energy description of such natural dynamical systems as turbulent water, earthquakes, and electrochemical dynamics in brain must be formulated in terms of fermions. We also believe that the methodology of the LEETs for chaotic dynamical systems may provide in the future a firm theoretical foundation for such high level concepts as the idea that the natural chaotic dynamical systems can be useful for information processing purposes understood broadly. 


\section*{Acknowledgments}

The author would like to thank Kang L. Wang for encouragement and discussions.



\end{document}